\documentstyle[twocolumn,epsf,rotate,floats,prb,aps]{revtex}

\draft

\input psfig

\begin{document}

\wideabs{
\title{Correlation between Spin Polarization and Magnetic Moment 
in Ferromagnetic Alloys}

\author{Tat-Sang Choy, Jian Chen, and Selman Hershfield} 
\address{{Department of Physics and National High Magnetic
               Field Laboratory,} \\
{University of Florida, 
 Gainesville, FL 32611} }

\maketitle

\begin{abstract}
The correlation between the 
magnetic moment in ferromagnetic alloys and the
tunneling spin polarization in 
ferromagnet-insulator-superconductor
tunneling experiments 
has been a mystery. 
The measured spin polarization 
for Fe, Co, Ni, and various Ni alloys 
is positive and roughly 
proportional to their magnetic moments, which can not
be explained by considering the net density of states.
Using a tight-binding coherent potential approximation 
(CPA) model,
we show that while the polarization of the
net density of states is not correlated with
the magnetic moment, the polarization of the 
density of states of {\it s} electrons
is correlated with the magnetic moment in
the same manner as observed by the tunneling experiments.
We also discuss the spin polarization measurements by
Andreev reflection experiments, some of which obtained different
results from the tunneling experiments and
our calculations. 
\end{abstract}

\pacs{PACS numbers: 75.70.-i, 75.10.Lp, 72.15.-v, 85.30.Mn}
}

\section{Introduction}

There has been renewed interest in spin-polarized transport
over the last decade.  This interest comes in part because of
a wide range of novel phenomena, e.g., the giant and colossal 
magnetoresistance,\cite{gmr,cmr}
spin-injection experiments,\cite{spininj} 
and spin-polarized tunneling 
experiments.\cite{tedrow2,julliere,maekawa,miyazaki,fif,zhang}
One of the most fundamental properties of spin polarized transport
in a ferromagnet is the polarization in the density of states
at the Fermi energy.  This polarization enters either directly
or indirectly into most transport calculations.  In particular,
since tunneling experiments measure the density of states,
they should provide a direct measure of this polarization.
In the case of ferromagnet-insulator-ferromagnet tunneling
experiments one measures the product of the spin 
polarizations.\cite{julliere}
However, in ferromagnet-insulator-superconductor tunneling
experiments where the density of states in the superconductor
is Zeeman split by a field in the plane of the film,
one can in principle measure directly the spin polarization
in the density of states.

Tedrow, Meservey, and collaborators carried out a series of 
ferromagnet-insulator-superconductor experiments 
on Fe, Co, Ni, and Ni alloys in the 1970's.
They found two surprising results. \cite{tedrow1}
First, the electron spin polarization for Fe, Co, and Ni 
is positive. 
Assuming the
tunneling conductance to be proportional to the total 
density of states at
the Fermi level,
band structure calculations
predict positive spin polarization for Fe and
negative spin polarization for Co and Ni,   
even though these
calculations have successfully explained
the Slater-Pauling curve \cite{sws2,friedel,crangle,dederichs} 
for the magnetic moments. 
Second, the spin polarization for Fe, Co, Ni, and the Ni 
alloys is roughly proportional to the magnetic moments. 
On the other hand, it is expected that only electrons near
the Fermi level participate in the tunneling process. Given the 
complicated
band structure of the transition metals, it is not clear why 
the electron spin polarization measured by tunneling, 
a property of the Fermi surface, and the magnetic moment,
a property of the Fermi whole sea, 
are related in such a simple way.

Recently, Soulen {\it et al.} \cite{soulen} and 
Upadhyay {\it et al.} \cite{upadhyay}
independently studied the spin polarization of 
ferromagnets with Andreev reflection point contact
experiments. 
The values of the spin polarization measured 
in these experiments are different from
those obtained in the tunneling experiments, although they
are in the same range. For example,
the spin polarization of Ni is 
43-46.5\% as measured by Soulen {\it et al.}
and 32\% as measured by Upadhyay {\it et al.}, 
while the tunneling spin
polarization of Ni measured by 
Moodera {\it et al.} is 33\%. \cite{fif}
The difference may be due to the experiments measuring
related but different quantities. One may have to take into 
account the different dependencies
on the density of states and the Fermi velocity
for two sets of experiments. \cite{mazin} 
The discrepancies in the experiments indicates that 
it is important for a theory to compare the trend instead of
particular values of the experiments. 
More recently, by using Andreev reflection techniques, Nadgorny {\it et al.} \cite{nadgorny}
measured the spin polarization of Ni$_x$Fe$_{1-x}$ alloys to be 
about 45\%,
roughly independent of the magnetic moment. 
Their results are different
from the tunneling measurements made by Tedrow and Meservey.

There have been a number of theoretical calculations to explain the
results of Tedrow and Meservey's tunneling experiments.
Many of these calculations concluded that the tunneling density
of states is dominated by only a fraction of the electrons.
In the work of Stearns \cite{stearns}, the relevant electronic states
are a $t_{2g}$-like band that is modeled by a parabolic dispersion
near the Fermi surface.
Hertz and Aoi \cite{hertz} 
concluded that tunneling measures the {\it s} density of states
as modified by many body effects due to the electron interaction
with spin waves.
Tsymbal and Pettifor \cite{tsymbal} 
studied the tunneling from Fe and Co by a tight-binding
model which has only
$ss\sigma$ bonds between the ferromagnet and the insulator.
They concluded that only {\it s} electron tunneling was sufficient to 
explain the experiments.
Recently, Nguyen-Manh {\it et al.} \cite{nguyen-manh}
performed a
self-consistent band structure calculation
of the Co/Al$_2$O$_3$ interface by a LMTO technique. 
Their calculation suggested that the interfacial cobalt {\it d} bands
spin polarize {\it s-p} bands in the barrier, resulting
in a positive tunneling spin polarization.
On the other hand, Mazin \cite{mazin} 
and Nadgorny {\it et al.} \cite{nadgorny} suggested that
spin polarization measured by both the tunneling and Andreev 
reflection experiments should be found from 
the polarization of the density of states times 
the Fermi velocity squared.
The values of the polarization obtained from the 
all of the above models are in the same range of the
experimental values; however, all these models are 
different and it is not
clear which model is closer to reality.

To understand if there is indeed a subset of the electronic 
states which can account for the
tunneling density of states, in this article we present a
microscopic calculation 
for {\em both} the magnetic moment and the various
density of states based on a self-consistent 
tight-binding coherent potential
approximation (CPA)
\cite{cpa} model.
In a range of alloys
we find that the {\it s} density of states follows the same trend
as the measured tunneling density of states.
This is the first microscopic calculation to see this correlation.
Furthermore, we are able to understand the correlation seen in
our calculation and to show that it is not universal, i.e.,
the tunneling density of states is not simply proportional
to the magnetic moment.  There may even be some alloys where they
are inversely correlated.

\section{Review of Experiments}

In this section, we review the series of tunneling experiments by
Tedrow and Meservey which show a correlation between the
spin polarization and the magnetic moment. We also
discuss the spin polarization measurements by 
Andreev reflection.

In the tunneling experiments of Tedrow and Meservey, 
spin polarized electrons tunnel from films made by alloying Ni 
and other 3{\it d} transition metals. 
In Fig. 1(a), the tunneling spin polarization for the Ni alloys
is plotted against the average number of valence electrons per atom,
which is changed by changing the composition of the alloys.
Figure 1(a) looks very similar to the 
Slater-Pauling curve shown in Fig.1(b), 
in which
the bulk magnetic moment of alloys are plotted against the 
average number number of valence electrons per atom.
In NiFe and NiMn, the spin polarization peak
is close to the magnetic moment peak.
In NiCu, NiCr, and NiTi, both the 
spin polarization and the magnetic moment 
decrease monotonically as impurity concentrations 
increase. The thresholds at which the spin polarization and the
magnetic moment drops to zero are close to each other.
These similarities suggest that
the spin polarization and the magnetic moment
are correlated.

\begin{figure}[htb]
\centerline{\psfig{file=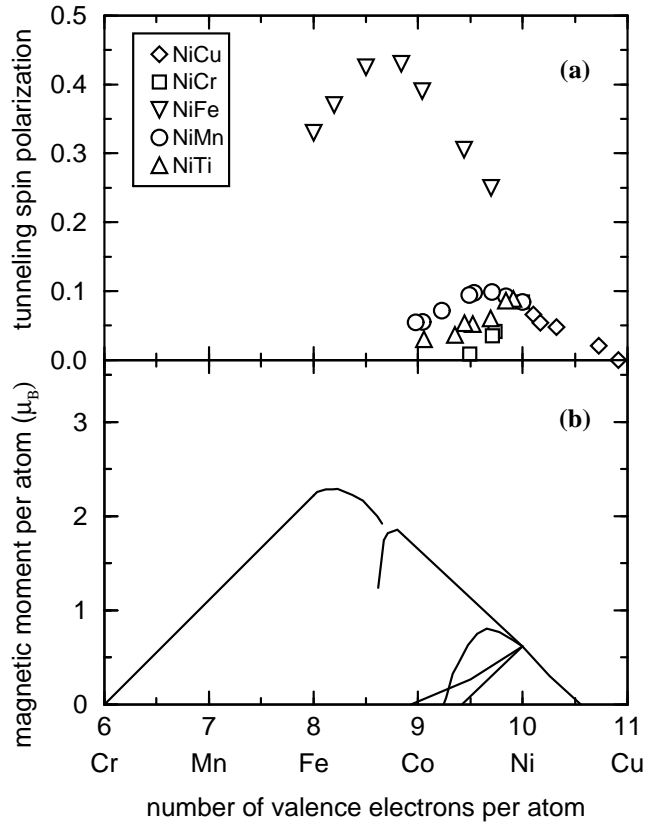,width=8.5cm,clip=}}
\caption{Experimental data of (a) tunneling spin polarization and
(b) bulk magnetic moment of 3{\it d} alloys versus
the average number of valence electrons per atom. 
The similarities between (a) and (b) suggest the tunneling
spin polarization and the magnetic moment are related.
}
\end{figure}

\begin{figure}[htb]
\centerline{\psfig{file=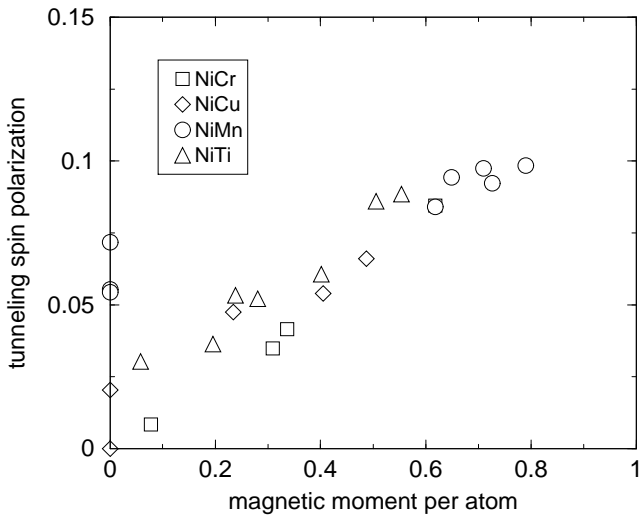,width=8.5cm,clip=}}
\caption{Experimental tunneling spin polarization \cite{tedrow1} of NiCr,
NiCu, NiMn, and NiTi versus the bulk magnetic moment of the corresponding
alloy. The data points for different alloys roughly line in the
same straight line. 
The commonly accepted experimental bulk magnetic moments of alloys 
are used as the x-coordinates.
Finite spin polarization is obtained
for zero magnetic moments in some samples.
This is probably due to the difference between
the bulk magnetic moments used on the graph and the surface
magnetic moments of the actual samples, which are not measured.
When the magnetic moment is zero, 
there should be no difference between the majority and minority
spins, the spin polarization is expected to
be zero.
}
\end{figure}

\begin{figure}[htb]
\centerline{\psfig{file=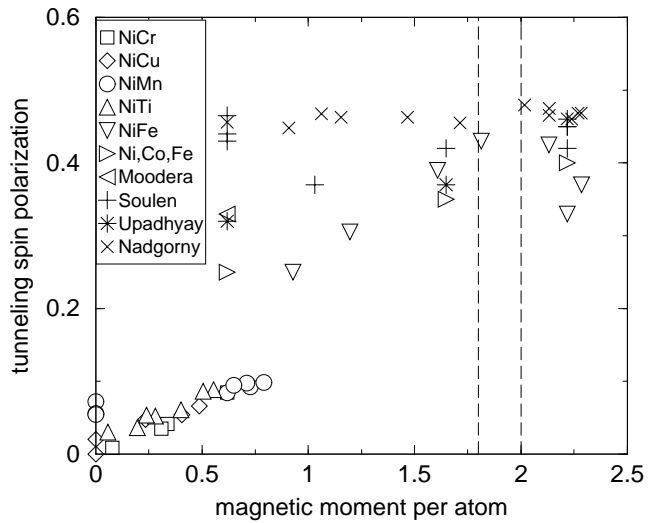,width=8.5cm,clip=}}
\caption{Spin polarization obtained by different experiments
versus the bulk magnetic moment of the corresponding
alloy. The open symbols are tunneling data and the rest are 
obtained from Andreev reflection experiments. 
The two dashed lines outline an area which roughly divides the graph 
into a fcc region (left) and
a bcc region (right).
}
\end{figure}

To see how the spin polarization relates to the magnetic moment,
we follow Tedrow and Meservey \cite{tedrow1}
and plot the spin polarization
against the magnetic moment for NiCr, NiCu, NiMn, and NiTi in
Fig. 2. The spin polarization data are taken from the
experiment of Tedrow and Meservey, and the
magnetic moment data are taken from the bulk measurements of 
alloys with the same compositions. 
Figure 2 shows a clear correlation
between the spin polarization and the magnetic
moment in these alloys.
The data points for different alloys
are roughly in a straight line passing through the origin.

Finite spin polarization is obtained for a few samples with
zero bulk magnetic moments. This is probably due to
the difference between the estimated bulk magnetic moments, which
are taken from experiments on alloys with the same composition, 
and the surface magnetic moments of the actual sample, which
are not measured in the tunneling experiments.
When the magnetic moment is zero, 
there should be no difference between the majority and minority
spins, the spin polarization is expected to
be zero.

In Fig. 3, we plotted the tunneling data shown
in Fig. 2 together with the tunneling measurements on other samples 
fabricated from different techniques by 
Tedrow and Meservey and by Moodera.
Also included are the spin polarization measurements using 
Andreev reflection
of Soulen {\it et al.}, Nadgorny {\it et al.}, 
and Upadhyay {\it et al.} 
The two dashed lines outline an area which roughly divides the graph 
into a fcc region (left) and
a bcc region (right). When a data point is close to this
area, the structure depends on the particular sample.
One can visually identify three regimes in the
figure: the lower left regime (fcc NiCr, NiCu, NiMn and NiTi), 
the middle regime (fcc NiFe with magnetic moment less than
about 1.8 $\mu_B$), 
and the right regime (bcc NiFe with magnetic moment greater than
about 2 $\mu_B$). 
There is a clear correlation between the 
spin polarization and the magnetic moment
in the lower left regime, which is the same as Fig. 2. 
In the middle regime, 
the tunneling data and the Andreev reflection 
data by Upadhyay {\it et al.}
only show a roughly increasing relation; 
on the other hand, the Andreev reflection data by 
Soulen {\it et al.} and Nadgorny {\it et al.} suggest that
the spin polarization is independent of the magnetic moment.
In right regime of Fig. 3, no clear 
correlation is seen between the spin polarization
and the magnetic moment. However, this regime corresponds 
to the peak area near Fe in Fig. 1, 
which clearly indicates a correlation between
the tunneling spin polarization and the
magnetic moment. 

\section{Model}

To see if there is actually a relationship between
the magnetic moment and 
the spin polarization of a subset of the electronic
states, 
we calculate both the magnetic moment and the spin polarization using
a tight-binding model. 
The band structure of an alloy
is calculated self-consistently within the coherent 
potential approximation.
The magnetic moments and 
the spin polarization of the density of states for different bands
at the Fermi level can then be obtained from the band structure. 
It is found that when the {\it p} and {\it d} bands are included, 
the spin polarization of the 
density of states is obviously inconsistent with the experiments. 
For example, negative spin polarization is 
obtained for Ni when the {\it d} bands are included.
Therefore, we present only the results of the spin polarization
of the {\it s} electrons. 
 
The alloys we consider are substitution alloys, 
in which some of the host atoms are replaced by impurity atoms
without changing the crystal structure.
The band structure of these alloys can be found
by the coherent potential approximation. \cite{cpa}

Before the band structure calculation of a magnetic 3{\it d} alloy 
can be carried out, one has
to know the splitting between the majority and
minority spins, which is related to the magnetic moment. 
The splitting can be found within our model from
the number of majority and minority electrons, 
which has to be found
from the band structure. Therefore, 
the band structure, 
the number of majority and minority electrons, 
and the splitting have to be solved self-consistently.

To calculate the alloy band 
structure, we consider a nine-band tight-binding Hamiltonian written as
\begin{equation}
H=\sum_{im\sigma} u_{im}^{\sigma}  c^{\dagger}_{im\sigma} c_{im\sigma} 
+ \sum_{\langle ij \rangle mn\sigma} 
t_{ijmn} c^{\dagger}_{im\sigma} c_{jn\sigma},  
\end{equation}
where $c^{\dagger}_{im\sigma}$($c_{im\sigma}$) is the 
creation (annihilation) operator
of a spin-$\sigma$ electron of orbital $m$ on the lattice site $i$, 
$u_{im}^{\sigma}$ is the on-site potential, $t_{ijmn}$ is the 
hoping energy between
neighbors. 
The band structure of the alloy described by this Hamiltonian
will be found using the coherent potential approximation.
Tight-binding parameters in the Slater-Koster two-center form
are obtained from 
fits to the local density approximation band structure.\cite{papa}
In principle all of the parameters
in the alloy are changed as one varies the alloy composition.
However, the 3{\it d} alloys are similar and the
differences in hoping and {\it s} and {\it p} on-site energies are
not as significant as the differences in the {\it d} on-site energies.
Therefore, 
we assume that the most important changes due to alloying
are contained in
the on-site energies,
$u_{i,d}^{\uparrow}$ and $u_{i,d}^{\downarrow}$, of the
spin-up (majority) and spin-down (minority) {\it d} electrons.
In other words, the alloy is assumed to have site-independent 
hoping parameters and {\it s} and {\it p} on-site energies 
the same as the host metal,
and site-dependent {\it d} on-site energies. 
This assumption will be justified later by the agreement with 
local density calculations of supercells in 
different impurity concentrations.

The major contributions to the on-site energies,
$u_{i,d}^{\uparrow}$ and  $u_{i,d}^{\downarrow}$ come from the 
atomic core potential plus the
Coulomb energy due to the opposite spin. 
In the itinerant electron model, it is more convenient
to work with the number of spin-up and
spin-down {\it d} electrons, 
$N_{i,d}^{\uparrow}$ and $N_{i,d}^{\downarrow}$,
at each site $i$. 
Therefore, we 
rewrite the parameters
in the form of 
\begin{eqnarray}
u_{i,d}^{\uparrow} &=& U^0_{i} + U^x_i N_{i,d}^{\downarrow}   \nonumber \\
u_{i,d}^{\downarrow} &=& U^0_{i} + U^x_i N_{i,d}^{\uparrow}, 
\end{eqnarray}
\noindent
where $U^0_i$ and $U^x_i$ are the on-site parameter 
and effective Coulomb
energy per pair at site $i$. 
Both $N_{i,d}^{\uparrow}$ and $N_{i,d}^{\downarrow}$ in Eq. (2) are
obtained from the band structure, which in turn depends
on $u_{i,d}^{\uparrow}$ and $u_{i,d}^{\downarrow}$. 
Thus, Eq. (2) has to be
solved self-consistently with the band structure calculated from
the Hamiltonian of Eq. (1).

All parameters for the host atoms can be found from
fits to the local density calculations of the pure metal. \cite{papa}
The only two parameters left, $U^0_{imp}$ and $U^x_{imp}$ 
of impurity atoms, are obtained from fits to
the local density calculations of supercells composed the 
two types of atoms.
For example, parameters for Fe as impurities in fcc
Ni host are obtained from fits to fcc Ni$_{3}$Fe 
band structures.
As a check, parameters obtained from the fit is used to 
calculate the tight-binding 
band structure of fcc Ni$_{31}$Fe, 
which agrees with the local density approximation band structure. 
This indicates that the 
model works well at least when impurity concentration is less 
than 25\%.
When the impurity concentration 
is large, or when the difference
in hoping between the host and the impurity is large, the
error is expected to increase.
Although this simplified model requires only two more parameters
in addition to the parameters of the host, 
it still produces correctly 
the Slater-Pauling curve for the magnetic moment \cite{choy}.

The three quantities which we will compare are
the magnetic moment per atom, the polarization of the
density of states
as measured by tunneling experiments, and the polarization of the 
calculated {\it s} density of states.  
The magnetic moment per atom, $\mu$, is determined from 
the number of electrons per atom in the majority 
spin orientation, $N^{\uparrow}$,
and the minority spin orientation, $N^{\downarrow}$, via
\begin{eqnarray}
\mu = \mu_B (N^{\uparrow} - N^{\downarrow}),
\end{eqnarray}
where $\mu_B$ is the Bohr magneton.
The {\it s} density of states of spin $\sigma$ at the 
Fermi level
at lattice site $i$ is defined as
\begin{equation}
n^{\sigma}_{i,s} = \int \frac{d^3 {\bf k}}{(2\pi )^3} 
\delta(E_F - E({\bf k})) \sum_{m=1}^9 | \langle i s \sigma|{\bf k} 
m \sigma \rangle |^2,
\end{equation}
where $|i s \sigma \rangle$ is the {\it s} orbital with spin $\sigma$
at site $i$ and $|{\bf k} m \sigma \rangle$ is the $m$-th eigenvector 
with a wavevector ${\bf k}$ and spin $\sigma$. 
The net {\it s} density of states is the weighted average of the 
{\it s} density of states at each lattice site:
\begin{equation}
n^{\sigma}_s = (1/N_{at}) \sum _i n^{\sigma}_{i,s} ,
\end{equation}
where $N_{at}$ is the number of atoms in the sample.
The density of states for spin $\sigma$ 
measured experimentally is
denoted by $n^{\sigma}$.
Using these definitions,
the polarization of the tunneling density of
states, $P$, and the calculated {\it s} density of 
states, $P_s$, are given by
\begin{eqnarray}
P &=& {{n^{\uparrow}-n^{\downarrow}} \over {n^{\uparrow}+n^{\downarrow}}}, 
\\
P_s &=& {{n_s^{\uparrow}-n_s^{\downarrow}} 
\over {n_s^{\uparrow}+n_s^{\downarrow}}}.
\end{eqnarray}

\section{Results}

As mentioned, when the density of states of the {\it p} and {\it d}
bands are included, 
the spin polarization is inconsistent with experiments. 
Therefore,
to compare with the experiments,
we plot the spin polarization of the {\it s} density of states, 
$P_s$,  against the calculated 
magnetic moment in the same graph as the experimental data.
In Fig. 4, we plot the calculated spin polarization of
NiCr and NiCu (filled symbols),
together with the experimental tunneling spin polarization, $P$, 
shown in Fig. 2 (unfilled symbols).
We choose to calculate NiCr and NiCu because they represent
different dependencies of the magnetic moment on
the average number of electrons. 
As seen in Fig. 1(b),
the magnetic moment of NiCr increases as 
the average number of electrons increases, while the
the magnetic moment of NiCu decreases as 
the average number of electrons increases.
For both alloys, the calculated spin polarization varies
from zero (non-magnetic) to about 27\% (pure Ni),
and has the same dependency on the magnetic moment.
When the magnetic moment is zero, the majority
and minority spin bands are the same, and
the spin polarization are also zero.

\begin{figure}[htb]
\centerline{\psfig{file=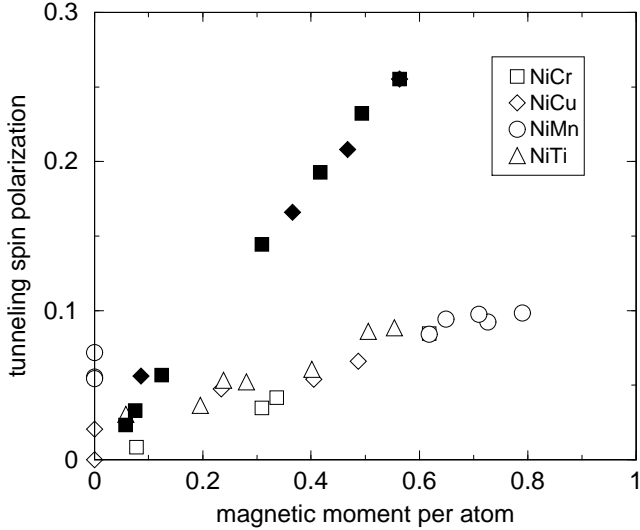,width=8.5cm,clip=}}
\caption{Calculated spin polarization of the {\it s} density of states
of NiCr (filled square)
and NiCu (filled diamond) alloys versus the magnetic moment plotted
with the experimental data (open symbols) shown
in fig 2.
The spin polarization measured for these series of films
are lower than the calculated values.
One reason is that
a growth technique that seems to reduce the spin polarization
was used in order to mix the elements. 
For example, the tunneling spin polarization measured 
for the Ni films in the series
are only about $8\%$, which is much lower than the values of
$27-33\%$ measured for the Ni films grown by better techniques.
}
\end{figure}

Both calculated and experimental results 
show that the spin polarization is 
positive and
roughly proportional to the magnetic moment. 
However, the experimental values
are much lower than the calculated ones. One reason may be that
in order to mix the elements,
the samples in this series of experiments were 
grown with technique which seems to reduce the 
spin polarization. \cite{tedrow1}
For example, the tunneling spin polarization measured for the Ni films
in this series is only about $8\%$, which is much lower than the 
measured values of $27-33\%$ for the Ni films grown by better
techniques.\cite{tedrow1,fif}

To compare with more experiments,
we plot in Fig. 5 the calculated spin polarization
of the {\it s} electrons for NiCr, NiCu, and fcc and 
bcc NiFe (filled symbols)
together with the experimental spin polarization
shown in Fig. 3 (unfilled symbols).
As explained above, the calculation for the lower left regime 
agrees with the tunneling experiment. 
The calculation for the middle regime (fcc NiFe)
agrees well with the tunneling experiments and the Andreev reflection
experiments by Upadhyay {\it et al.}, but not with the 
Andreev reflection experiments of 
Soulen {\it et al.} and Nadgorny {\it et al.}
In the right regime (bcc NiFe), the 
calculated results are significantly
higher than the measured values.

\begin{figure}[htb]
\centerline{\psfig{file=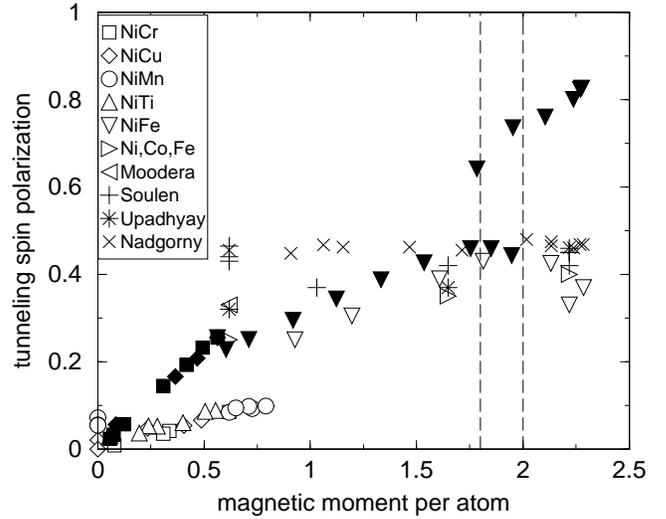,width=8.5cm,clip=}}
\caption{Calculated spin polarization of the {\it s} density of states of
 NiCr (filled square),
 NiCu (filled diamond), and NiFe (filled downward triangle)
 alloys versus the magnetic moment plotted with
 the
 experimental data (open symbols) shown in Fig 3. 
}
\end{figure}

Thus far we have presented calculations of the bulk density of states. 
However, experiments have suggested that the 
spin-polarized tunneling electrons
originate from the first two to three layers 
at the surface. \cite{tedrow2,tedrow1}
To estimate how the surface affects the spin polarization, 
we study the variation of the {\it s} density of states as a function
of depth from the surface. 
The band structure of 
bcc (100) Fe,
hcp (001) Co, and fcc (111) Ni slabs of
18 atomic layers thick
are calculated by using fixed tight-binding parameters. 
This calculation only serves as a crude estimate 
of the surface effects.
Self-consistent iterations are not used in
the band structure calculation. Effects such as the
change in surface magnetic moment and surface roughness 
are also neglected.
We study the spin polarization of the 
cumulative {\it s} density of states
as a function of the number of layers.
The cumulative {\it s} density of states of spin-$\sigma$ 
in the first $l$ layers from the surface is defined as 
${\overline n^{\sigma}_{l,s}} = \sum_{j=1}^{l} n^{\sigma}_{j,s}$,
where $n^{\sigma}_{j,s}$ is the {\it s} density of states in the
$j$-th layer.
The spin polarization of the cumulative {\it s} density of states
for the first $l$ layers from the surface is given by
\begin{eqnarray}
{\overline P_{l,s}} &=& 
{{{\overline n_{l,s}^{\uparrow}}-{\overline n_{l,s}^{\downarrow}}} 
\over {{\overline n_{l,s}^{\uparrow}}+{\overline n_{l,s}^{\downarrow}}}}.
\end{eqnarray}
This quantity is related to the tunneling spin polarization because
it takes into account the electrons contributed from the 
first few layers.
However, this is only an approximation to the 
tunneling spin polarization 
because the contribution from
different layers may not be uniformly weighted.

\begin{figure}[htb]
\centerline{\psfig{file=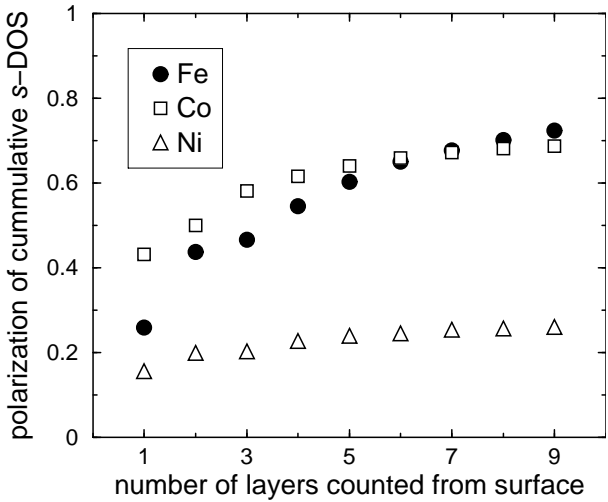,width=8cm,clip=}}
\caption{Calculated spin polarization of the cumulative {\it s} density of states,
$\overline{P_{l,s}}$, as a function of the number layers counted from the surface.
We studied the spin polarization of the cumulative {\it s} density of states
 because tunneling may be sensitive to the first few layers.
As shown in the graph, 
the spin polarization of the {\it s} density of states 
at the bcc Fe surface is significantly
lower than that in the bulk. This reduction is
less important in the hcp and fcc metals.
Therefore, 
our estimation of the tunneling spin polarization using the bulk
spin polarization is good for the fcc metals, while
it is too large in the case of the bcc metals,
as shown in fig 5.
}
\end{figure}

As shown in Fig. 6, there is a reduction of the spin
polarization near the surface. The reduction is more important 
in bcc metals than in fcc metals.
For example, the spin polarization of bcc Fe surface is 
less than one third
of the bulk value, while the spin polarization of the fcc Ni surface
is about two thirds of the bulk value.
The degree of reduction in the spin polarization
depends mainly on the structure of film. Therefore, as seen
in Fig. 5, our calculation for the spin polarization of the bcc alloys
using the bulk spin polarization is 
significantly higher than the experimental value, 
which may be related to
the surface spin polarization. 
For the same reason, the calculated 
spin polarization of the fcc alloys agrees
well with the experiments.

One may now ask the question: why should the polarization of the
{\it s} density of 
states, $P_s$, and the magnetic moment per atom, $\mu$,
be related?
In Fig. 7 we have plotted 
the {\it s} density of states, $n_{i,s}^{\sigma}$,
for the two spin orientations, $\sigma$, 
at the different sites, $i$, in fcc Ni rich
alloys.  The x-axis is 
the total number of electrons, $N_i^{\sigma}$, at 
site $i$ with spin $\sigma$.  As evident from the figure, for a range
of compositions and different alloys, 
all the points lie on the same curve.
The solid and dashed lines are obtained from pure Ni 
by varying the Fermi energy in
the spin-up and spin-down bands respectively.
This curve contains the key to understanding 
why the {\it s} density of
states and the magnetic moment are related.  
In the range of $3.0 < N^{\sigma}_i < 5.5$, 
$n^{\sigma}_{s,i}$ is an {\it increasing} 
function of $N^{\sigma}_i$ for both spins. 
Since $n^{\sigma}_{s}$ increases as 
$N^{\sigma}$ increases, 
it follows that the spin polarization,
$P_s \propto (n^{\uparrow}_{s} -  n^{\downarrow}_{s})$, increases 
with increasing magnetic moment,
$\mu = \mu_B (N^{\uparrow} - N^{\downarrow})$.

\begin{figure}[htb]
\centerline{\psfig{file=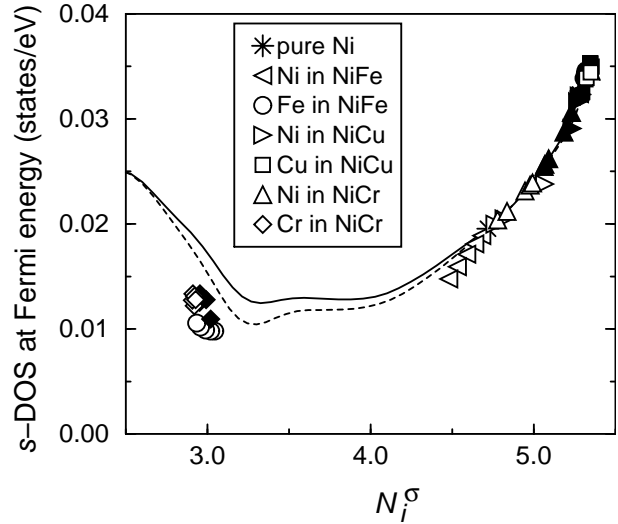,width=8cm,clip=}}
\caption{The {\it s} density of states $n^{\sigma}_{s,i}$ at the Fermi energy 
for spin-up (filled symbols) and spin-down (unfilled symbols) 
channels versus
the number of spin-up and spin-down valence electrons,
 $N^{\uparrow}_i$ and $N^{\downarrow}_i$, at site $i$ ($i=$ Ni, Fe, Cu, \ldots)
in fcc alloys.
The solid (dashed) curve is obtained by
varying the Fermi energy in the spin-up
(spin-down) density of states of pure Ni. 
In the range of interest, ($N^{\sigma}_i>3$) 
$n^{\sigma}_{s,i}$ is an increasing function of $N^{\sigma}_i$. 
It follows that
$P_s \propto (n^{\uparrow}_{s} -  n^{\downarrow}_{s})$ increases as
$\mu = \mu_B (N^{\uparrow} - N^{\downarrow})$ 
increases, explaining the correlation in Fig. 1-3.}
\end{figure}

It is important to note
that $P_s$ 
is only roughly proportional to $\mu$ 
because the curve in Fig. 7 is quite different
from a straight line even in the range of $3.0 < N^{\sigma}_i < 5.5$. 
Furthermore, this is not a universal relation because the curve
shown in Fig. 7 is for the fcc Ni rich alloys.  Other kinds of alloys
would presumably produce a different curve, possibly even a decreasing
instead of an increasing curve.
For example, applying the same analysis to bcc alloys 
such as FeCr shows that
the {\it s} density of states is still 
an increasing function of $N^{\sigma}_i$ 
in the range
of interest.  However, as shown in Fig. 8, a much different curve
is obtained for the bcc alloys. In bcc alloys, 
the s density of states
has sudden jumps in the energy range of interest. 
One of the jumps is reflected
in Fig. 8 by the curves near $N_i^{\sigma} = 4.5$. 
The jump causes
a sudden increase in the spin polarization of 
the {\it s} density of states 
in bcc FeCr with low Fe concentration,\cite{choy} 
which has not been studied experimentally.

\begin{figure}[htb]
\centerline{\psfig{file=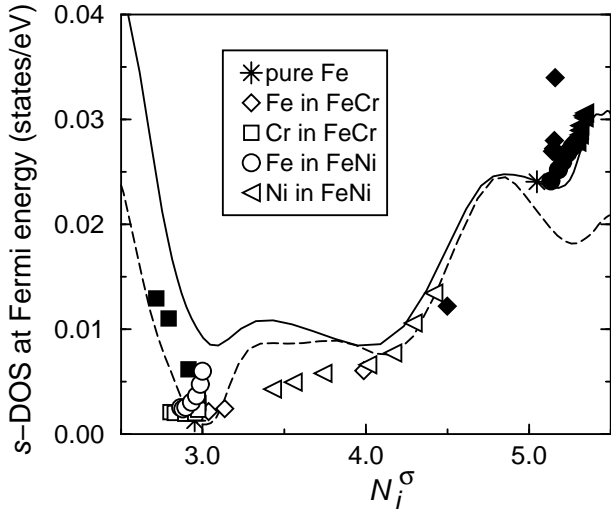,width=8cm,clip=}}
\caption{The {\it s} density of states $n^{\sigma}_{s,i}$ at the Fermi energy 
for spin-up (filled symbols) and spin-down (unfilled symbols) 
channels versus
the number of spin-up and spin-down valence electrons,
 $N^{\uparrow}_i$ and $N^{\downarrow}_i$, at site $i$ ($i=$ Ni, Fe, or Cr)
in bcc alloys.
The solid (dashed) curve is obtained by
varying the Fermi energy in the spin-up
(spin-down) density of states of pure Fe. 
In this case,
$n^{\sigma}_{s,i}$ is still an increasing function of $N^{\sigma}_i$
in most region; however, the function is more complicated than that
of the fcc alloys.
}
\end{figure}

While this explains the correlation, it still leaves open the
question of why all the Ni rich alloys fall on a single curve in Fig. 7
and why that curve is increasing.
To understand this we examine more closely the band structure and 
in particular the density of states for the {\it s} and {\it d} bands.
The {\it d} density of states is responsible for the magnetic moment, 
while as
argued above the {\it s} density of states is 
responsible for the tunneling.
The {\it s}  and {\it d} density of states are related
by {\it s-d} hybridization.

As an example, we have plotted in Fig. 9 the {\it s} and
{\it d} density of states near the Fermi level 
for (a) the majority band
and (b) the minority band
at the Ni sites of 
pure Ni, Ni$_{0.8}$Fe$_{0.2}$, and Ni$_{0.6}$Fe$_{0.4}$. 
We have shifted the energy such that within the same spin channel, 
the integrated density of states from the bottom of 
the bands to 0eV ({\it not} the Fermi level) 
are the same for every alloy. 
The Fermi energies in these plots are indicated by the vertical lines.
The {\it s} density of states fall on the same curve and the 
{\it d} peaks have the same 
energy. The only difference among the alloys are the Fermi levels.
Thus, the primary effect of alloying at the Ni sites is to shift the 
{\it s} and {\it d} bands {\it together} relative to the Fermi level.
We see from Fig. 9 that the {\it s} density of states $n^{\sigma}_s$
of the fcc alloys is an increasing function of the Fermi energy in
the range where the Fermi level lies. 
On the other hand, 
$N^{\sigma}$, the integrated density of states up to the Fermi level,
also increases with the Fermi energy.
Therefore $n^{\sigma}_s$ is an increasing function of
$N^{\sigma}$ as shown in Fig. 7.

\begin{figure}[htb]
\centerline{\psfig{file=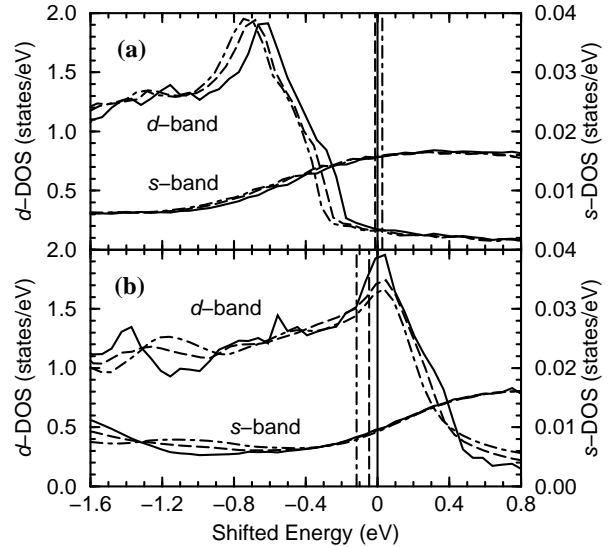,width=8cm,clip=}}
\caption{The (a) spin-up and (b) spin-down {\it d} and 
{\it s} density of states
at the Ni 
sites in pure Ni (solid lines), Ni$_{0.8}$Fe$_{0.2}$ (dashed lines) and 
Ni$_{0.6}$Fe$_{0.4}$ (dot-dashed lines) 
versus the shifted energies. 
The energy of each spin channel for each alloy is shifted such that the 
number of states at the Ni sites below 0eV
is the same as pure Ni in the corresponding spin channel. 
The {\it d} peaks roughly coincide and 
the {\it s} density of states collapse onto a single curve. 
Therefore the primary effect of alloying is to shift
the bands together relative to the Fermi level.
This explains why data for
different alloys and spins in Fig. 7 are on the same curve.
The positive slope in Fig. 7 can be explained by
noting that the Fermi levels lie in the range
where the {\it s} density of states increases as a function of energy and
hence the number of electrons $N^{\sigma}_i$.
}
\end{figure}

It should be noted that the shifting of the {\it d} bands is due 
to energy considerations as in the itinerant electron theory.
It results in the change of the magnetic moments, $\mu$. 
On the other hand, the shifting of the {\it s} band is due to its 
hybridization with the shifted {\it d} bands. It results in the
change of the spin polarization of the {\it s} density of states,
$P_s$.

\section{Conclusion}

In this paper, a microscopic calculation for both the magnetic moment
and the spin polarization of subsets of the density of states of the
3{\it d} alloys are studied.
By interpreting the tunneling spin polarization as 
the spin polarization 
of the {\it s} density of states, 
the trends in the tunneling experiments of 
Tedrow and Meservey are obtained.
The correlation between the 
magnetic moment and the density of states 
are understood by showing 
that the {\it s} density of states for both spin orientations
is an increasing function of the number of electrons 
and by showing that the primary effect of alloying is to shift 
the bands together relative to the Fermi level.
The correlation between the magnetic moment and the
polarization of the {\it s} density
of states is not universal.  
All of this work
supports the picture that the tunneling current is dominated
by {\it s} electrons. 

While we have explained the correlation between the 
spin polarization and the magnetic moment,
there are still a number of open questions.
Within this model,
the tunneling current is assumed to be dominated by {\it s} electrons,
but the mechanism behind it is not clear.
Tsymbal and Pettifor\cite{tsymbal}
suggested that the hoping from the {\it d} bands of
the ferromagnet to the {\it s} band of the barrier is essentially zero. 
In 3{\it d} metals, the {\it s-d} hoping is normally 
a few times smaller
than the {\it s-s} hoping, so it is reasonable to 
assume a similar ratio for the
hoping at the metal-insulator interface.
However, we note that since the 
{\it d} density of states is about two orders
of magnitude higher than that of the 
{\it s} density of states at the Fermi level, 
the {\it s-d} hoping has to be much smaller than 
a tenth of the {\it s-s} hoping to explain the dominance of the 
tunneling current
by the {\it s} electrons. Therefore, 
it is unclear if such a requirement is physical. 

In another model, suggested
by Nguyen-Manh {\it et al.},\cite{nguyen-manh}
the {\it s} band in the insulator 
is spin-polarized by the {\it d} band of
the ferromagnet due to hybridization, 
causing a positive spin polarization in the {\it s} current. The model
predicts that in the insulator there is very small but spin polarized 
density of states 
at the Fermi level.

On the other hand, a different model was suggested by
Mazin\cite{mazin} and Nadgorny {\it et al.}.\cite{nadgorny}
They argued that the 
current, both in tunneling and Andreev reflection experiments,
is proportional to the density of states times the 
Fermi velocity squared. The
low(high) density of states of the {\it s}({\it d}) electrons 
are therefore compensated by
their high(low) Fermi velocities. Thus, both the {\it s} and the
{\it d} electrons are important to
the tunneling current.
The spin polarization they calculated is roughly 
independent of the magnetic moment, much like the 
Andreev reflection data the
group obtained.
At this point, it is unclear which of the above 
models gives a better physical picture.

Another open question is whether the spin polarization measured by
the tunneling experiments and
the Andreev reflection point contact experiments
are the same. Mazin\cite{mazin} and Nadgorny {\it et el.}\cite{nadgorny}
argued that the two are the same. 
However, the situation here is more 
complicated. There are even disagreements among the results
obtained from the Andreev reflection experiments. 
The results
obtained by Soulen {\it et al.}\cite{soulen}
and Nadgorny {\it et al.}\cite{nadgorny} are not the same as those obtained
by Upadhyay {\it et al.}\cite{upadhyay} 
It is unclear whether this is due to the differences in the
sample preparation 
or the method of data analysis.

There are also qualitative differences between 
the prediction of different 
models. While our calculations show that the spin
polarization increases with magnetic moments, 
the calculations by Mazin\cite{mazin} and 
Nadgorny {\it et al.}\cite{nadgorny} show that
the spin polarization is independent of the magnetic moments.
However, the spin polarization can remain constant
at most in a certain regime. 
In regimes such as Ni$_{1-x}$Cr$_{x}$ with $x > 0.15$, 
 Ni$_{1-x}$Cu$_{x}$ with $x > 0.55$, or the invar regime
near Ni$_{0.36}$Fe$_{0.64}$, the magnetic moment drops to zero.
When the magnetic moment is zero
the spin polarization is expected be zero because there is 
no difference between the majority and minority spins. 
It would be helpful to compare
experiments and theories in these regimes.

\end{document}